# COMPARATIVE STUDY OF VARIOUS VOIP APPLICATIONS IN 802.11A WIRELESS NETWORK SCENARIO

Sutanu Ghosh
Faculty of Electronics and Communication Engineering,
Dr. Sudhir Chandra Sur Degree Engineering College

*ABSTRACT*

*Today, Voice over Wireless Local Area Network (VOWLAN) is the most accepted Internet application. There are a large number of literatures regarding the performance of various WLAN networks. Most of them focus on simulations and modeling, but there are also some experiments with real networks. This paper explains the comparison of performance of two different VOIP (Voice over Internet Protocol) applications over the same IEEE 802.11a wireless network. Radio link standard 802.11a have maximum transmission rate of 54Mbps. First protocol is session initiation protocol (SIP) and second is H.323 protocol. First one has an agent called SIP proxy. Second have a gateway reflects the characteristics of a Switched Circuit Network (SCN). With this comparison we have required to obtain a better understanding of wireless network suitability for voice communication in IP network.*

## 1.INTRODUCTION

Presently, the corporate world and the large community choose packet-switched networks to carry voice traffic instead of the traditional circuit-switched TDM network. There have two protocols in this domain for a long time, namely H.323 and SIP as specified earlier. The protocol H.323 is recommended by most important regulatory body in telecommunications ITU-T [9, 13], whereas SIP is developed by IETF (a body which regulates the development of Internet) defined in RFC 2543 [10]. This paper describes comparative analysis of their characteristics in particular 802.11a Wireless LAN scenario. There is same kind of view in a literature [1] for these two protocols, but it is significantly well defined for 802.11a network and I want to produce some graphical representations of different parameters through the comparison. Those parameters are depending upon VOIP initiator and receiver. Parameters are likely Session establishment time, total number of bytes sent and received etc.

## 2. WIRELESS LOCAL AREA NETWORK

IEEE 802.11 Wireless LAN has become a common domain of network technology. It performs the functions like a traditional LAN except the wireless interface. As compared to wired LAN; actual throughput in WLAN is depending upon the product and set-up. The Throughput can be affected by number of users, latency on the wireless part of the WLAN, transmission rate, transmitted power, multipath propagation, type of WLAN used etc. Effective throughput is the total transported data payload for a fixed transmission time [2]. It uses the unlicensed ISM frequency band (Industrial, Scientific and Medical). The ISM band has three frequency ranges: 902-928 MHz, 2400-2483.5 MHz and 5725-5850 MHz. WLAN physical layers (PHYs) provide various data rates with different modulation and coding techniques. There have two standard MAC specifications at WLAN: the contention-based Distributed Coordination Function (DCF) and the polling-based optional Point Coordination Function (PCF). DCF uses carrier sense multiple access technology with collision avoidance (CSMA/CA) mechanism and PCF uses centralized polling based access mechanism. The collision avoidance function uses different amount of delays to minimize the risk of a collision. The network architecture consists one or more access point and client devices. The IEEE 802.11 standard provides two architectures - Infrastructure and Adhoc mode. These modes are used to build wireless networks. The Infrastructure mode must have single or more Access Points (AP) to establish the communication process between mobile stations within a same cell. It can be possible to extend the network through the interconnection between several Access points. Wireless nodes can move between different Access Points within the same network for this mode. Adhoc mode does not contain access points. The stations can communicate directly.

In this research, I can see that 802.11a network is standard because this radio link has better throughput, less amount of interference and deployment cost is very low. But it has a better response in a shorter range only. It supports 5.7 GHz radio frequency band with the bandwidth of 20 MHz. The data rate is maximum 54 Mbps with the OFDM modulation technique. Actually OFDM technique provides 8 different modes of data rate - 6, 9, 12, 18, 24, 36, 48, and 54 Mbit/s [3]. The frequency band 5.7 GHz is an unlicensed national information infrastructure band. So, it is used to support these 8 PHY modes with different data transmission rates. The PHY is the interface between MAC and wireless medium. It transmits and receives data frames over the shared wireless medium. There have an exchange of a frame in between MAC and PHY is under the control of the Physical Layer Convergence Procedure (PLCP) sub layer. Now, MAC Protocol Data Unit (MPDU) consists of MAC header, variable-length information frame body, and frame check sequence (FCS). During the transmission process, a PLCP preamble and a PLCP header are added with the MPDU to form a PLCP Protocol Data Unit (PPDU). PPDU format of the IEEE 802.11a PHY contains PLCP preamble, PLCP header, MPDU (part of MAC), tail bits and pad bits, if require.

**DCF:** Carrier sensing in DCF is done through the physical and virtual mechanism. A station senses the medium to listen if other station is transmitting, in the physical mechanism. The virtual mechanism of carrier-sense is done by distributing reservation information technique with RTS/CTS Exchange. If the medium is free then station waits for a specified time called distributed inter-frame space (DIFS), and then station executes data transmission. If the medium is busy then current station executes a back-off algorithm within a contention window (CW). Backoff is activated if the medium were found busy, when sensing for first time. If the backoff is initiated, then station waits till the present transmission is over. At the end of the present transmission, station waits for a DIFS. If the medium was detected to be idle for a DIFS, then the station takes an additional backoff wait before transmission. The backoff timer has taken as some randomly chosen value. If the medium is completely free throughout the backoff interval and expires the timer, the frame is sent. If the medium is busy during the backoff interval, the timer is fixed at its current value. Station waits for the time the medium to become free, waits for another DIFS, and waits again for another backoff, decreasing the timer value. This process continues till the end of the backoff timer and the station sends the data. When a station accepts the DATA frame, it waits for SIFS (Short IFS) interval and sends an ACK to the sender.

The channel has been sensed idle for a period equal to,
DIFS + Backoff Time ;

Where, DIFS=SIFS + 2* Slot Time
Backoff Time= Random() * Slot Time,
Random() is a pseudorandom number drawn from a uniform distribution over the interval [0, CW] within the range $CW_{min} \leq CW \leq CW_{max}$;
$CW_{min}$, $CW_{max}$ and Slot Time depend on the physical characteristics.

Table 1. Contention window and slot time for IEEE 802.11a [14]

|  | $CW_{min}$ | CWmax | Slot Time |
|---|---|---|---|
| IEEE 802.11 a | 16 | 1023 | 9 μsec |

**PCF:** Time is segregated into superframes. Superframe consists of the contention period (CP) where DCF is useful and a contention-free period (CFP) where PCF is useful. The CFP is executed by beacon frame transmitted by the PC using DCF. CFP may change from one superframe to other. PC polls stations in a Round-Robin Fashion. The polling method has done according to the polling list. If there have no pending transmission, then response has a null frame (that has no payload). If the CFP ends before the execution of total method, then polling list is restarted at the very next station in the following CFP succession. PCF IFS (PIFS) is used by the Access Point to gain access over the medium before all other stations.

## 3. VOIP NETWORK

It is the process to define the voice call transmission over the internet protocol. The basic technique of transmission is to break a voice stream into the number of small packets, then compressed and finally sent them towards the destination through various routes. At the receiver end the packets are reassembled, decompressed and converted back into a voice stream by various hardware and software. The system consist pair of encoder-decoder and an IP transport network. There have a requirement of audio codec to encode the speech to allow the transmission over an IP network. This codec is a set of computer program to implement an algorithm that compresses/decompresses digital audio data according to the given audio file format or the streaming media audio format. Now, we should concentrate on the different types of VOIP applications.

**H.323** is gateway-gatekeeper based VOIP application which defines protocol, procedure and different network component to support a good multimedia communication capability over packet based network. Packet based network may contain IP network, enterprise network, internet packet exchange based local area network or metropolitan area networks/wide area networks. The application is valuable for any kind of network like date based, audio based or audio-video based etc. The main components of H.323 are Terminals, Multipoint controller, Multipoint Control Units (MCUs), Gateways, Gatekeeper and Peer Elements. Terminals are simply Telephones, IVR devices, Video phones, Voicemail Systems etc. The multipoint controller (MC) provides the interaction between three or more endpoints like MCU in a multipoint conference. There have a multipoint processor (MP) is used to collect audio, video or data streams and distributes them to endpoints participating nodes. MCU consists of an MC and one or more MPs. If it provides the centralized multipoint conferences, then MCU consists MC, audio, video, and data MP. Gateway contains Media Gateway Controller (MGC) and Media Gateway (MG). MGC controls the call signaling and other non-media-related functions whereas MG controls the media related part. This Media Gateway and Media gateway controller may co-exist or exist separately. Gateway is not required unless the switch circuit network is used.

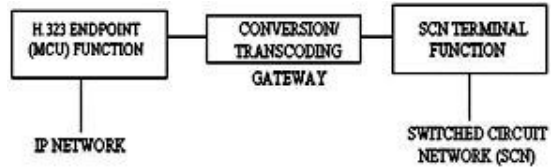

Figure 1. Basic model H.323 Gateway

Gatekeepers are optional component, used to control admission and address resolution. It is logically separated from other H.323 network components. It have required to provide the set of services such as addressing, authorization, bandwidth management, billing, charging, accounting, authentication of terminals and gateways. Peer elements are co-located with Gatekeeper. They can transfer the addressing information and participate in call authorization between administrative domains. The elements may gather address information to reduce the volume of routing information passes through the network. They support in call authorization/authentication directly between two administrative domains or via the clearing house.

**Session Initiation Protocol (SIP)** is used to control the initiation, modification and termination of interactive multimedia sessions. The multimedia sessions may be established as audio or video calls among two or more parties or subscriber, chat sessions (audio or video chat) or game sessions. Instant messaging, presence and event notifications can be defined by SIP extensions. Basically, SIP is a text-based protocol which is similar to HTTP [8] and Simple Mail Transfer Protocol (SMTP). SIP contains 4 ideal components - User Agent, Registrar, Redirect server and Proxy server [7]. The end point of SIP is a user agent. Gateway is the connector and translator between packet and circuit switch networks. The server is used to keep a track of user information called Registrar. Redirect server is another server, which is used to inform the devices when they must contact external domain to perform a specific function. SIP proxy server is involved with the connection and session establishment. There have five signaling messages [15] to control the SIP protocol. Those are REGISTER, INVITE, BYE, CANCEL and ACK. The first message is used to register a user agent with SIP Proxy. Session initiation and termination is possible through the INVITE and BYE message respectively. Connection can be discarded via the message of CANCEL. The message ACK is used for consistent information exchange for invitations.

**VOIP packet format [15]:**

RTP Header:  RTP is the real time transport protocol, used for the transmission of audio or video stream. This parameter is one of the important things for the VOIP application. RTP supports the samples to be reconstructed in the proper sequence of order and gives us a technique for the measurement of delay and jitter. The size of the header is 12 in bytes.

UDP Header: UDP is a best effort service to insert 8 octets and routes the data to the actual destination port. It is simply connectionless and does not support any sequence information or the guarantee of delivery of information.

IP Header:  IP inserts 20 bytes for VOIP packet and is responsible for delivering the data to the correct destination. It is another simple connectionless and does not guarantee of delivery or the sequence of order.

Data: Payload size is varying from 20 bytes to 160 bytes.
**VOIP Codec:**  The conversion process of analog waveform to the digital form is carried out by a codec. Various types codec present in real time application – GSM 6.10, G.711, G.729, G.723.1 etc. Codec samples the waveform at regular intervals and generates a value for each sample. These samples are typically taken 8,000 times/sec. These individual values are gathered for a fixed period of time to create a frame of data. Generally, a sample period of 20 ms is very common. Some codec use longer sample periods, such as 30 ms employed by G.723.1. Others use shorter periods, such as 10 ms employed by G.729.

## 4. RELATED WORK

There have several works under the same issue but the domain is different. In this section, I will review the earlier few works of comparison between two protocols by outlining their main ideas and nextly describe my work in the specific domain & compare these two protocols using a combination of analysis, measurements and simulation of the protocol operations simultaneously.

### 4.1. A Comparison of SIP and H.323 for Internet Telephony

By Schulzrinne and Rosenberg, 1998 [4].

This paper compares SIP and H.323 versions 1 and 2. Basically it is a qualitative and descriptive comparison of the two protocols. The main issues of comparisons are complexity, extensibility, scalability and services. It can be observed that the real time protocol controls the media exchange. The conclusion is that SIP provides same types of services, compare to H.323, but supports less complexity, better extensibility and scalability.

### 4.2. Comparison of H.323 and SIP for IP Telephony Signaling

By Dalgic and Fang, 1999 [5].

The paper gives a brief idea about 2 protocols and creates a comparative literature of those protocols on various classes. Actually, they put the description of call control services and the details of signaling exchanged for two protocols. The main conclusions are in terms of functionality and services that can be supported, H.323v3 and SIP are very similar. The Authors have found that SIP has better flexibility and H.323 has better compatibility with its various versions and enhanced interoperability with the PSTN.

### 4.3. A Comparison of H.323v4 and SIP

By Nortel Networks, 2000 [6].

This work compares SIP and H.323 version 4 (most popular), on the basis of complexity, extensibility, scalability, resource utilization, resource management, services and measures the performance of them in a wireless environment. The main idea is to establish call control protocol produces a significant advantage over the other in terms of the various categories. The report considers that UDP service is used for both of the protocols. The experimental results are based on call flows of each protocol for many services, focusing the particular case where a SIP or H.323 entity communicates with a UMTS 2000 entity and the group of work provides a few numerical results. The Authors mainly highlight the issues regarding the interaction of each protocol with the UMTS 2000 architecture. Finally, the authors conclude and recommend the SIP as their own choice of control protocol. SIP provides long standing effects which are associated and involved with time to market, multi-party service flexibility and extensibility, ease of interoperability and complexity of development.

## 4.4. Comparison of SIP and H.323 Protocols

By Ilija Basicevic, Miroslav Popovic and Dragan Kukolj, 2008 [1].

It is the most recent work in this comparative issue. They have highlighted on those characteristics that are most important for successful operation of VoIP network. It includes the characteristics of syntax to functions of infrastructure elements and support for third party call control and other communications applications. Security features are also analyzed.
SIP has HTTP based text syntax and the extensions are mostly in xml. In case of H.323, syntax based on ASN.1.
Conclusion of the paper is that H.323 appeared earlier than SIP, but SIP has better adaptation to the internet environment. SIP has also greater standardization activity related to applications.

## 5. SIMULATION ENVIRONMENT

Before this work, I have done a same kind of comparative analysis with SIP application. In this research issue consider a WLAN network that have 9 nodes. Three of them are mobile nodes. There have one extra node 10 (which is used as SIP proxy in SIP network and same is used to present general interface node for H.323 network). Node 10 is used to perform a generalized function for both of the VOIP network (H.323 and SIP). Node 1, 2, 3, 4, 5 and node 6, 7, 8, 9 are connected to wireless network1 and wireless network2 respectively. Node 10 is locally connected to node 5 and 6.

Table 2. Wireless Network Channel set up (bitwise) nodes

| Network | Listenable channel | Listening channel |
|---|---|---|
| Wireless Network 1 | 0100 | 0100 |
| Wireless Network 2 | 0010 | 0010 |
| Wireless Network 3 | 0001 | 0001 |

Table 3. Link Application set up between the nodes

| Link Established between the nodes | Application |
|---|---|
| 4 to 5, 3 to 7, 1 to 9, 2 to 8, 5 to7 | VOIP |
| 4 to 6, 3 to 8 | FTP |
| 1 to 9 | CBR |

I have chosen the mobile nodes as host 1, 7 and 8. Host1 have a slow movement through the waypoint 1, 2 and 3 and a fast movement at waypoint 4. Host 7 and 8 have a slow movement in compare to 1. I have simulated the scenario in the Qualnet [11] environment for two times for different applications and collect the result to compare the performance. The total simulation time have taken as 134 seconds. Here the flags are used to present the path of the movement of mobile nodes.

**Global Setting:** There have 3 different channels have taken for this scenario. Two of them are used for Wireless LAN and last one for backbone. The power, gain and frequency etc are defined as follows:

*Transmission power*: 39 dBm for various data rate, *Directional antenna gain*: 15dB,
*Path Loss Model*: Two Ray model

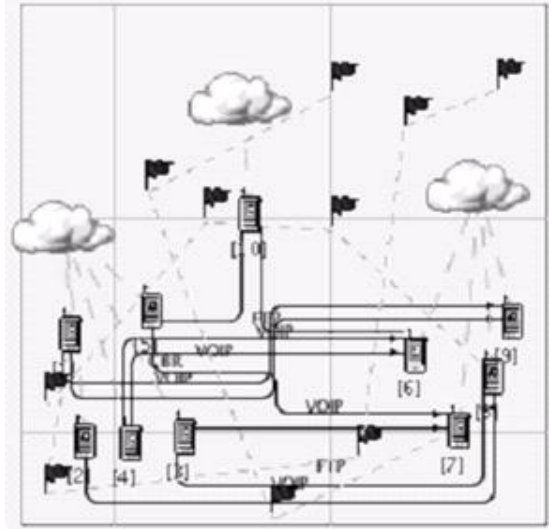

Figure 2. Simple Qualnet Simulation Scenario model

## 6. COLLECTIVE DATA ANALYSIS

This scenario has lots of application, but I want to describe only VOIP applications for my study of comparison. After the simulation, I have collected, analyzed and compared the data for both initiator & receiver of two different applications. For performance analysis I have taken only 5 parameters for this research issue –

i) Session Establishment time, ii) Total Bytes Sent, iii) Total Bytes Received, iv) RTP average end to end delay and v) Overall Throughput.

Because these five parameters are more convenient for the calculation of throughput of my scenario.

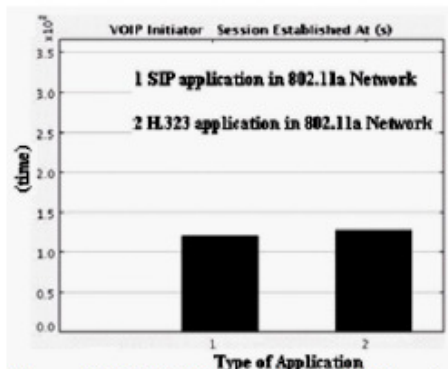
Figure 3. VOIP initiator session establishment time (in sec)

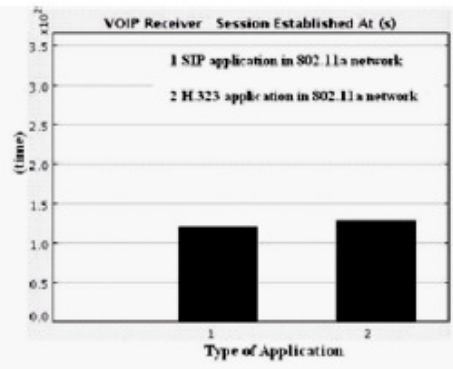
Figure 4. VOIP Receiver session establishment time (in sec)

The initiator session establishment time for SIP application is 125 seconds, whereas H.323 has required 131 seconds to establish the session. The receiver session establishment time for SIP and H.323 application is exactly same as the initiator session establishment time.

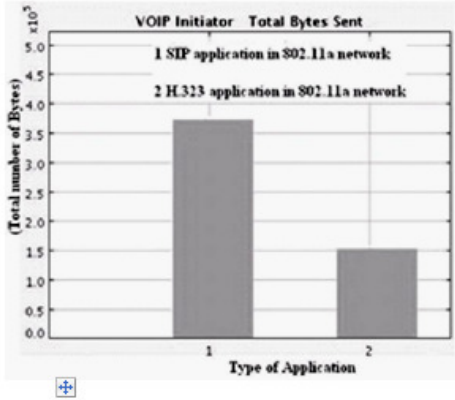
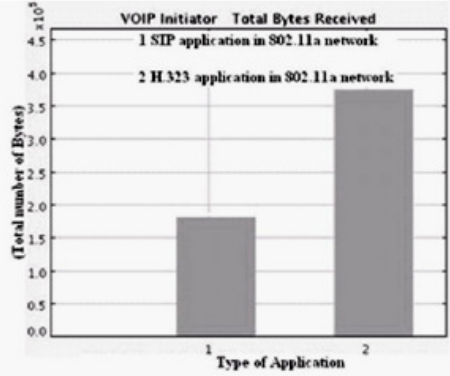

Figure 5. VOIP Initiator total Bytes Sent

Figure 6. VOIP Initiator total Bytes received

In the fig. 5 shows that SIP application has sent more number of Bytes compare to H.323 for a same amount of time. This result is just opposite of fig.6, which is a comparison between maximum number of Bytes received in H.323 and SIP initiator.

It is easy to identify from fig.5 and fig.8 that the total Bytes sent from the initiator and total Bytes received in the receiver is exactly same.

The comparison between fig.6 and fig.7 proves that total Bytes sent from the receiver and total Bytes received in initiator is same.

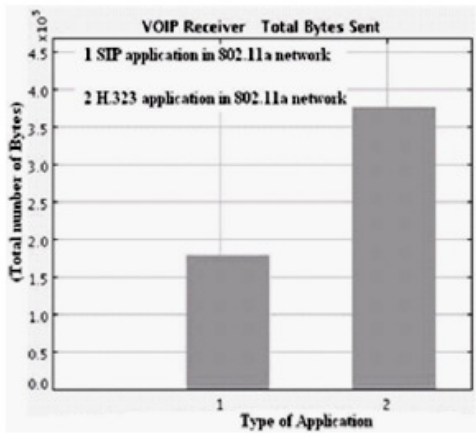
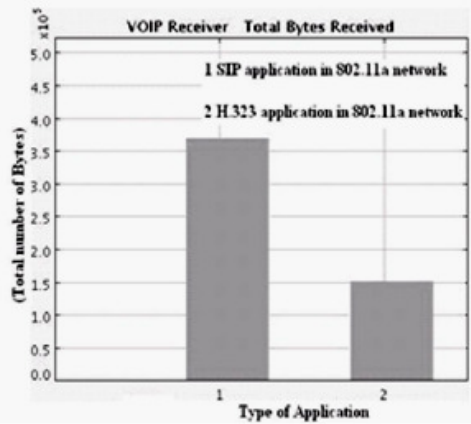

Figure 7. VOIP Receiver total Bytes Sent

Figure 8. VOIP Receiver total Bytes received

It is very easy to understand from the last four figures that when SIP initiates the process then it has greater number Bytes sent and received over the H.323 application. The opposite case can be possible in the same scenario when H.323 receiver initiates the process.

International Journal of Mobile Network Communications & Telematics ( IJMNCT) Vol. 3, No.5, October 2013

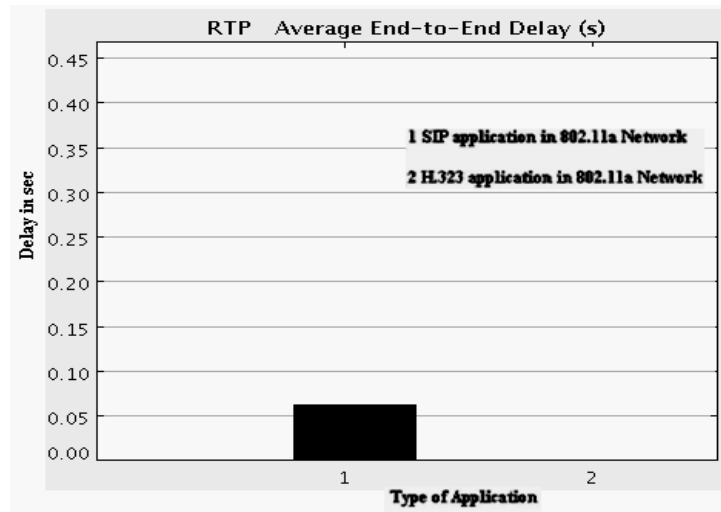

Figure 9. RTP average end to end delay comparison

This comparison based on a specific codec. The above figure depicts that the delay amount for SIP application 0.05 sec. But delay amount is not fixed. It is different for different codec in VOIP applications.

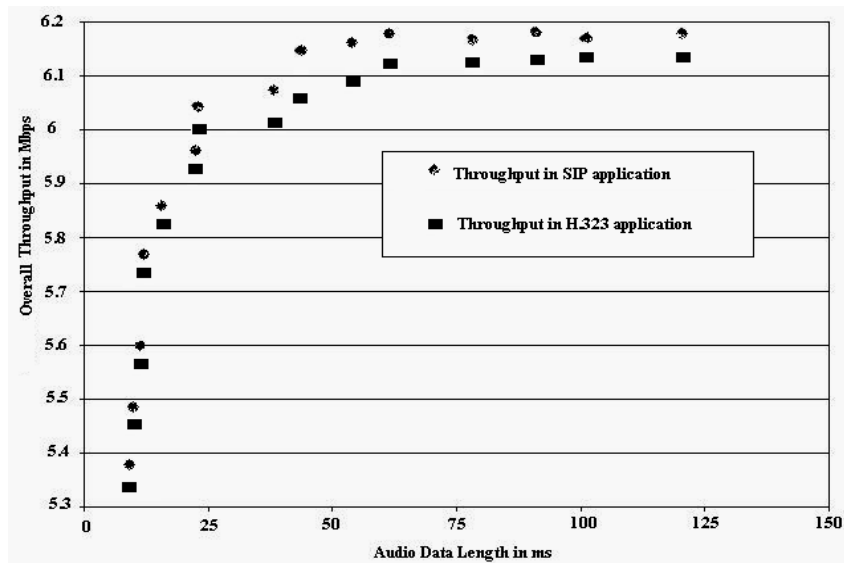

Figure 10. Overall throughput comparison for two different VOIP applications with respect to the duration of Audio length

The above result gives a good comparison between two applications for the issue of throughput calculation. The overall throughput is slightly better for SIP application in same time duration. The maximum throughput is 6.182 Mbps at the duration of 123 ms and 6.122 Mbps at same duration for SIP and H.323 application respectively.

## 7. CONCLUSIONS AND FUTURE WORK

It has been analyzed from different parameters that VOIP initiator talking time is greater for

37

802.11a. Simple conclusion can be drawn here that SIP has same number of packet drops in jitter buffer with H.323 for 802.11a network but the establishment time and session initiation is better & faster for SIP application. So, my research conclusion is same with Ref. [5] for specific domain of wireless environment 802.11a.

SIP supports better than H.323 in busiest area of communication. Here, I have used same codec (G.711) in most of the time of my work. I have taken the standard value of sample duration for the codec (G.711 have sample duration of 15ms). The main objective of this experiment has to gain some idea regarding the relationship between the distance on the communications and the quality aspects. This has produced a useful result for the next research issue where the environment has taken in the larger area inside the office. This kind of wireless scenario is a better way of understanding of wireless network suitability for voice communication in IP network. This issue is purely based on 802.11a network; it can be analyzed for other wireless environment. And the performance analysis of this comparative literature can be extended using the other network metrics.

Note : Fig. is the short form of figure.